\documentclass[10pt,twoside,BCOR7mm,DIV15,headinclude,footexclude,cleardoubleempty,idxtotoc]{scrartcl}

\usepackage[english]{babel}
\usepackage{graphicx}
\usepackage{hyperref}
\usepackage{scrpage2}
\usepackage{hyperref}
\usepackage{ifthen}

\makeatletter
\renewcommand{\@biblabel}[1]{}
\renewcommand{\@cite}[2]{%
{#1\ifthenelse{\boolean{@tempswa}}{,#2}{}}}
\makeatother

\hypersetup{breaklinks=true
,colorlinks=true,linkcolor=black,urlcolor=blue
,citecolor=black}

\ofoot{\thepage}
\ifoot{}

\setheadsepline{1pt}

\setkomafont{pagehead}{\normalfont\sffamily}
\setkomafont{pagenumber}{\normalfont\rmfamily}

\usepackage{booktabs}
\usepackage{amsmath}
\usepackage{amssymb}
\usepackage{multicol}
\usepackage{float}

\makeatletter
\newcommand{\listofcontributions}{\@starttoc{con}}

\newcommand{\l@contribution} {\@dottedtocline{1}{1.5em}{2.3em}}
\makeatother

\newenvironment{contribution}{
\setcounter{section}{0}
\setcounter{figure}{0}
\setcounter{table}{0}
\begin{flushleft}
{\em Clumping in Hot Star Winds \\
W.-R.\ Hamann, A.\ Feldmeier \& L.\ Oskinova, eds.\\
Potsdam: Univ.-Verl., 2007 \\
URN: http://nbn-resolving.de/urn:nbn:de:kobv:517-opus-13981
} 
\end{flushleft}
}{
\newpage
\lehead{}
\rohead{}
}

\begin{document}

\setlength{\baselineskip}{2.5ex}

\begin{contribution}

\newcommand\sun{\odot}
\newcommand\arcdegr{\mbox{$^\circ$}} 
\newcommand\arcmin{\mbox{$^\prime$}}

\newcommand\aj{{AJ}}
\newcommand\araa{{ARA\&A}}
\newcommand\apj{{ApJ}}
\newcommand\apjl{{ApJ}}
\newcommand\apjs{{ApJS}}
\newcommand\apss{{Ap\&SS}}
\newcommand\aap{{A\&A}}
\newcommand\aapr{{A\&A~Rev.}}
\newcommand\aaps{{A\&AS}}
\newcommand\baas{{BAAS}}
\newcommand\mnras{{MNRAS}}
\newcommand\pasp{{PASP}}
\newcommand\pasj{{PASJ}}
\newcommand\ssr{{Space~Sci.~Rev.}}
\newcommand\nat{{Nature}}
\newcommand\iaucirc{{IAU~Circ.}}
\newcommand\aplett{{Astrophys.~Lett.}}
\newcommand\apspr{{Astrophys.~Space~Phys.~Res.}}

\lehead{A.\ Reimer}

\rohead{Clumping effects on non-thermal particle spectra}

\begin{center}
{\LARGE \bf Clumping effects on non-thermal particle spectra in massive star systems}\\
\medskip

{\it\bf A. Reimer}\\

{\it W.W. Hansen Experimental Physics Laboratory \& Kavli Institute for Particle Astrophysics \& Cosmology, Stanford University,
452 Lomita Mall, Stanford, CA 94305, USA}

\begin{abstract}
Observational evidence exists that winds of massive stars are clumped. Many massive star systems are known as
non-thermal particle production sites, as indicated by their synchrotron emission in the radio band. As a
consequence they are also considered as candidate sites for non-thermal high-energy photon production up to
gamma-ray energies. 
The present work considers the effects of wind clumpiness expected on the emitting relativistic particle spectrum
in colliding wind systems, built up from the pool of thermal wind particles through diffusive particle acceleration,
and taking into account inverse Compton and synchrotron losses. In comparison to a homogeneous wind, a clumpy wind
causes flux variations of the emitting particle spectrum when the clump enters the wind collision region. It is found that the spectral features associated with this variability moves temporally from low to high energy bands with the time shift between
any two spectral bands being dependent on clump size, filling factor, and the energy-dependence of particle energy gains and losses.
\end{abstract}
\end{center}

\begin{multicols}{2}

\section{Introduction}

Evidence for particle acceleration to 
relativistic energies mediated by the supersonic 
winds of massive, 
hot 
stars comes from the observation of non-thermal radio emission 
(e.g., \cite{Abbott1986}).
This has been interpreted by synchrotron emission on the basis of the
measured spectra (much steeper than the canonical value $\alpha_r\sim +0.6$,
$F_\nu\propto \nu^{\alpha_r}$) and high 
brightness temperatures of $\sim 10^{6-7}$K, far exceeding $\sim 10^4$K
expected from free-free emission from a steady-state isothermal radially 
symmetric wind (\cite{Wright1975}). Those particles has been suggested 
to be accelerated
either in shocks caused by the instability of radiatively driven winds
(\cite{White1985}), in the shocked wind collision region of multiple systems
(e.g., \cite{Eichler1993})
or in the termination shock (\cite{Voelk1982}).

Through a statistical study the presence of non-thermal radio emission has been linked
to the binarity status of the stellar systems (\cite{Dougherty2000}),
which is in support of a scenario where particles being predominantly accelerated
at the forward and reverse shocks from the colliding supersonic 
winds from massive stars.
Therefore, in this work I am considering the non-thermal particle spectra from being produced
in the collision region of a typical long-period massive binary system.

By now ample evidence has accumulated that hint towards the clumpiness of WR- and O-star winds:
non-converging mass loss rates from various methods (thermal radio, $H_\alpha$, UV-lines, X-ray and IR diagnostics, etc.), polarization and photometric variability, stochastically variable substructures in lines (e.g., \cite{Eversberg1998}), the observation of non-thermal
radio emission even in some short period binaries (e.g., CygOB2\#8A: \cite{Blomme2005}) where self-absorption in the
optically thick winds should prevent the visibility of a non-thermal component, etc.
Still open, however, remains a qualitative assessment about the filling factors, typical clump sizes, clumping startification, etc.

\section{Emitting electron spectrum}

This work considers, for the first time, the effect of clumpiness in the colliding winds of binary systems on the resulting
non-thermal electron spectrum in the collision region at a given orbital phase, assuming that the particles to be accelerated 
stem from the pool of thermal 
wind particles. It requires a fully time-dependent treatment of particle injection, acceleration and losses.

For this purpose I describe diffusive particle acceleration and losses (radiatively, and via escape from the collision region with rate $T_0^{-1}$)
that governs the non-thermal emitting electron spectrum, by the kinetic equation:
\begin{equation}
\frac{\partial}{\partial t}(N(E,t)) + \frac{\partial}{\partial E}(\dot E\, N(E,t)) + \frac{N(E,t)}{T_{0}} = Q(E,t)
\end{equation}
where ${\dot E=aE-\dot E_{\rm loss}}$, the radiative energy loss rate ${\dot E_{\rm loss}}=(b_{\rm syn}+b_{\rm IC})E^2$
from synchrotron and inverse Compton losses in the Thomson regime, the acceleration rate ${a =\frac{V_{\rm OB}^2 (c_r-1)}{3c_r\kappa_{\rm a}}}$ (\cite{Reimer2006}) with ${c_r}$ the shock compression ratio, and $\kappa_{\rm a}$ the diffusion coefficient, ${T_0}=\frac{r_0}{V}$ the escape time scale for a constant post-shock flow velocity $V$, and the size of the acceleration region, assumed cylinder-shaped for simplicity, with
$r_0=\kappa_{\rm a}/V$. 
In this picture the clumpiness of the wind translates into a variable injection rate $Q(E,t)$, 
while for a homogenuous wind
$Q(E,t)\propto \delta(E-E_0)$.

In the following I consider the typical setting of a colliding wind region (see e.g., \cite{Eichler1993})
at a given orbital phase with a clumpy wind of volume filling factor $f_{\rm vol}$ defined as the 
total volume occupied by the clumps with respect to the total wind volume, 
$f_{\rm vol}=V_{\rm clump}/V_{\rm tot}$. The simplified picture of a constant particle density within the
clump and no particles in the interclump medium used here is sufficient to expose the basic properties of 
variability in a non-thermal component from a clumpy wind.
Furthermore, $f_{\rm vol}$ and $V_{\rm OB}$ shall be constant sufficiently close to the shock region. 

$Q(E,t)=Q_0 \delta(E-E_0) H(t-t_0) H(t_v-t)$ describes the particle injection into the shock region
for a clump of size $l_{\rm clump}=V_{\rm OB}(t_v-t_0)$. The solution is analytic, and shown
in Fig.~\ref{AReimer:Fig1} for $t_v\rightarrow\infty$. With the spectral index $s$ of the emitting electron 
spectrum, $\propto E^{-s}$,
determined by the acceleration rate and escape, the exponential built-up of the electron spectrum up to a 
maximum energy $E_c=a/(b_{\rm syn}+b_{\rm IC})$ (see Fig.~\ref{AReimer:Fig2}, solid line) is governed by both, 
energy gains and losses. Steady-state is reached on hours time scale in the typical settings
of the colliding wind region in long-period binary systems.

The interclump phase is described by $Q(E,t) \propto E^{-s}\delta(t-t_v)$ as the source term in Eq.~1. For $a>0$ in this phase
the solution of Eq.~1 implies a continued acceleration with the maximum electron energy increasing further exponentially
(see Fig.~\ref{AReimer:Fig2}, dashed line): the particle acceleration appears as a multi- (here: two)-stage process.
For $a=0$ further acceleration ceases and the emitting electron spectrum declines on typically 
hour times scales in the astrophysical environment considered here. Variations of the acceleration rate $a$ may be plausible
considering that possible variations of the shock conditions (e.g. compression ratio, etc.) as the clumpy winds collide re not unexpected.

Combining injection and interclump phases
and taking into account past clump injections into the shock results
in a temporally evolving electron spectrum in the wind collision region. An example for $l_{\rm clump}=10^{12}$cm is 
shown in Fig.~\ref{AReimer:Fig3}. Clearly visible are the spectral features from injection phases, and they propagate
from low to high energies as time progresses. Thus the flux variations from clumpy particle injections into the
acceleration process in the shock at low energies are anticipated to precede the 
flux variations at high energies. Ultimately, the underlying physical reason for this time shift is the unavoidable
energy dependence of the radiative particle energy losses, and their gains in the acceleration process.
In addition to the clump size and volume filling factor, the expected time shift is therefore dependent on
the electron energy losses and gains.

\begin{figure}[H]
\begin{center}
\includegraphics[width=\columnwidth]{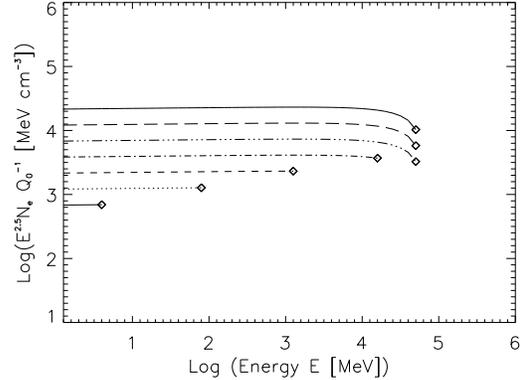}
\caption{Injection phase: Resulting emitting electron spectrum at times 1, 3, 5, 7, 10, 15 and 20 ksec 
(correspondingly, lower to upper lines)
after $t_0=0$ in response to injecting particles into the acceleration process starting at $t_0$.
Parameters used for all figures presented in this work:
bolometric luminosity $L_{\rm OB}=10^6L_\sun$ and effective temperature $T_{\rm eff}=45000$K of the star closer
to the stagnation point, terminal wind velocities $V_{\rm WR}=V_{\rm OB}=2000$km/s, mass loss rates of the OB, $\dot M_{\rm OB}=10^{-6}M_\sun$/yr, and WR-star, $\dot M_{\rm WR}=10^{-5}M_\sun$/yr, constant magnetic field in the shock region B=1~G,
a stellar separation of $D=5\cdot 10^{14}$cm, postshock flow velocity $V=0.6V_{\rm OB}$, diffusion coefficient 
$\kappa_a=10^{19}{\rm cm}^2$/s, shock compression ratio $c_r=4$ and $E_0=1$MeV. Steady-state is reached on hours time scale.
\label{AReimer:Fig1}}
\end{center}
\end{figure}

\section{Discussion and Outlook}

In this work consideration has been given to the effects on the non-thermal electron spectra,
as produced in a diffusive shock acceleration process from the pool of thermal wind particles, when the 
massive star winds are clumped as they collide. In this case, the electron injection into the shock
varies with time, dependent on the volume filling factor and typical clump size. Consequently, flux variations
are expected in non-thermal particle and photon spectra, in addition to the thermal emission component.
I have shown that those flux variations are expected to appear time shifted when comparing various
spectral bands, with the low energy flux variations preceding those at high energies. This clumping signature
in non-thermal spectra probes wind clumping in the very vicinity of the collision region, and
may possibly show up in multifrequency observations at energy bands that are dominated by non-thermal emission,
provided sufficiently dense data sampling. 

A metallicity dependence of clumping in non-thermal spectra may be probed by comparing clumping signatures
from non-thermal ion spectra, appearing eventually in the gamma-ray domain only through $\pi^0$-decay photon 
emission, with those from
non-thermal electron spectra. For this purpose we plan to apply the presented prescription to
clumpy ion injections as well.

The presented scenario describes a simplified clumping picture, which nonetheless proves to be
sufficient to expose the basic properties of a variable particle injection into the acceleration process
in colliding wind regions. This general prescription allows to also account for more complex 
clumping configurations (e.g. density structure within clump and/or interclump medium, irregular nested clump
sizes on various scales, etc.), by modifying the injection term $Q(E,t)$ accordingly.

\begin{figure}[H]
\begin{center}
\includegraphics[width=\columnwidth]{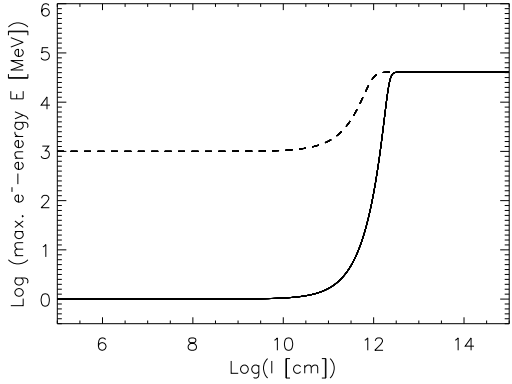}
\caption{Exponential increase of maximum energy as a function of time with $l=V(t-t_0)$. During the clump injection phase
the maximum energy reaches a level
$\frac{1}{E_{\rm max,1}(t)} = \frac{1}{E_c}-\left(\frac{1}{E_c}-\frac{1}{E_0}\right)\exp\left(-a\frac{l_{\rm clump}(t)}{V}\right)$ (solid line).
During the interclump phase the maximum particle energy increases further if $a>0$ as
$\frac{1}{E_{\rm max,2}(t)} = \frac{1}{E_c}+\left(\frac{1}{E_{\rm max,1}}-\frac{1}{E_c}\right)\exp\left(-a\frac{l_{\rm interclump}(t)}{V}\right)$ (dashed line). $E_{\rm max,1}=10^3$~MeV and the rate of energy gain $a=$const at all phases is used here for demonstration.
\label{AReimer:Fig2}}
\end{center}
\end{figure}

\begin{figure}[H]
\begin{center}
\includegraphics[width=\columnwidth]{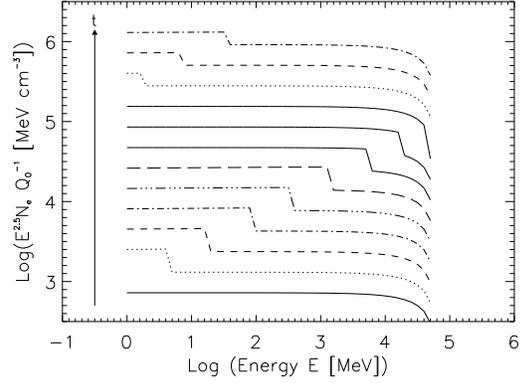}
\caption{Temporal evolution of the total emitting electron spectrum in the wind collision region with clumpy particle injection
and clump size $l_{\rm clump}=10^{12}$cm. A temporal step size of $\Delta t=1$ks is used.
All spectra are separated artificially in normalization for optimal visibility. $a=$const at all phases is used 
here as an example.
\label{AReimer:Fig3}}
\end{center}
\end{figure}



\end{multicols}

\end{contribution}


\end{document}